\documentclass[aps,prd,reprint,amsmath,amssymb,nofootinbib,floatfix,10pt]{revtex4-2}

\usepackage{bbold}
\usepackage{graphicx}
\usepackage[font=footnotesize]{caption}
\usepackage[font=footnotesize]{subcaption}
\usepackage[colorlinks=true,linkcolor=blue]{hyperref}
\usepackage[capitalise]{cleveref}

\newcommand{\C}{\mathbb{C}}
\newcommand{\eps}{\varepsilon}
\newcommand{\id}{\mathbb{1}}
\newcommand{\ii}{\mathrm{i}}
\newcommand{\SL}{\mathrm{SL}}
\newcommand{\SU}{\mathrm{SU}}
\newcommand{\tr}{\mathrm{tr}}

\graphicspath{{./figures/}}

\begin{document}

    \title{Finite-density equation of state of hot QCD using the complex Langevin equation}

    \author{Michael Mandl}
    \email{michael.mandl@uni-graz.at}
    \affiliation{Institute of Physics, NAWI Graz, University of Graz, Universitätsplatz 5, 8010 Graz, Austria}
    \author{Dénes Sexty}
    \email{denes.sexty@uni-graz.at}
    \affiliation{Institute of Physics, NAWI Graz, University of Graz, Universitätsplatz 5, 8010 Graz, Austria}
        \author{Daniel Unterhuber}
    \email{daniel.unterhuber@uni-graz.at}
    \affiliation{Institute of Physics, NAWI Graz, University of Graz, Universitätsplatz 5, 8010 Graz, Austria}

\begin{abstract}	
	We present the results of continuum-extrapolated lattice simulations of quantum chromodynamics (QCD) above the crossover temperature and for unprecedentedly high baryon densities at the physical point, employing the complex Langevin equation. In particular, we determine the QCD equation of state by computing the baryon density as well as the pressure as functions of the baryon chemical potential and the temperature. Potential issues with wrong convergence of complex Langevin dynamics are under control and we indeed find agreement with previous lattice studies working at smaller chemical potentials, as well as with perturbative hard-thermal-loop calculations at high temperatures.
\end{abstract}	

\maketitle

	\emph{Introduction.}
	Quantum chromodynamics (QCD) is the theory describing the strong interaction between quarks and gluons. Besides governing the confinement of these fundamental constituents into bound states like protons and neutrons, QCD encodes the physics of various other phenomena of contemporary interest. For instance, the strong interaction plays a major role in collision experiments of heavy ions, e.g., at RHIC in Brookhaven or the LHC at CERN, in compact stellar objects such as neutron stars, and in the early stages of the Universe like the nucleosynthesis era. For the theoretical description of these systems, a proper understanding of QCD under extreme conditions such as high temperature and baryon density is of fundamental importance \cite{BP25r}.

By now, the temperature axis of the QCD phase diagram is well understood owing to \emph{ab initio} lattice simulations, which have predicted, for instance, that the high-temperature transition of QCD is really an analytic crossover \cite{AEF06}. For nonzero baryon chemical potentials, on the other hand, lattice simulations face the infamous sign problem \cite{TW05}, severely restricting the range of baryon densities one can reliably access. The most prominent approaches based on reweighting, Taylor expansion, or imaginary chemical potential are currently limited to baryon chemical potentials about two to four times the temperature \cite{ADE05,BEF12,BDH17,DGS17,BDH19,GKK20,BFG21,BGK22,
BFG22,BGK22_2,BFG23,BCG23,BFG24,ABF25}. The presence of the sign problem thus calls for alternative approaches not based on importance sampling. While various ideas for such alternatives exist, none of them has been able to fully circumvent the sign problem of QCD for arbitrary baryon densities. 

One particularly appealing solution to the sign problem is the complex Langevin approach \cite{Par83,Kla83}, which is based on the complexification of the underlying field degrees of freedom and their stochastic evolution in an artificial time dimension -- a generalization of stochastic quantization \cite{PW81} to complex variables; see, e.g., \cite{BRL21} for a review. Since the introduction of the gauge cooling technique for stabilizing simulations \cite{SSS13,NNS16_2}, its application to finite-density QCD has shown promising results \cite{Sex14,Sex14_2,ASS14,FKS15,ASS17,NNS18,Sex19,KS19,IMN20,SSS20_2,
AJZ20r,AJZ22u,HS24,TAI25} and even produced a full phase diagram in the heavy-dense limit \cite{AAJ16}. Moreover, important progress has been made in the theoretical understanding of complex Langevin dynamics and its limitations -- in particular, the notorious wrong-convergence problem --, resulting in the development of several diagnostic tools that can be used to assess the correctness of simulation results \cite{ASS10,AJS11,NNS16,SSS19,SSS20,SSS24,MSS25,JK25_1,Man26}, further establishing the reliability of the method. Up to this date, however, complex Langevin simulations of QCD were restricted to unphysical pion masses.

In this work, we present the first complex Langevin results for QCD at the physical point, including a continuum extrapolation. We reach unprecedentedly high baryon chemical potentials $\mu_B$ as compared to approaches relying on importance sampling, but are restricted to temperatures $T$ above the crossover. Here, we discuss our main results for the QCD equation of state (see, e.g., \cite{Phi13r}) in the $(\mu_B,T)$ plane including the baryon density as well as the pressure, while a more detailed analysis, covering also additional thermodynamic quantities, will be published elsewhere \cite{MSUu}.

	\emph{Lattice setup.}
	We have performed large-scale lattice simulations using complex Langevin dynamics. In the context of QCD, this entails the complexification of the usual $\SU(3)$ gauge links to noncompact $\SL(3,\C)$-valued ones. The stochastic evolution of the complexified gauge links then proceeds via the discrete update equation
\begin{align}\label{eq:langevin_update}
	\begin{aligned}
		\tilde{U}_{x\mu}^{(n)} = \exp\left[\ii\left(\eps K_{x\mu}[U^{(n)}]+\sqrt{\alpha\eps}\eta_{x\mu}^{(n)}\right)\right]U_{x\mu}^{(n)}\;, \\
		U_{x\mu}^{(n+1)} = \exp\left[\ii\left(\eps\gamma\tilde{K}_{x\mu}^{(n)} + \sqrt{\alpha\eps}\eta_{x\mu}^{(n)}\right)\right]U_{x\mu}^{(n)}\;,
	\end{aligned}
\end{align}
with $\alpha = 2-3\eps$, $\gamma=\frac{1}{2}+\frac{\eps}{4}$,
\begin{equation}\label{eq:langevin_update_aux}
	K_{x\mu}[U] = \sum_a\lambda_aK_{x\mu a}[U]\;, 
	\quad
	\eta_{x\mu}^{(n)} = \sum_a\lambda_a\eta_{x\mu a}^{(n)}\;,
\end{equation}
and 
\begin{equation}
	\tilde{K}_{x\mu}^{(n)} = K_{x\mu}[U^{(n)}] + K_{x\mu}[\tilde{U}^{(n)}]\;.
\end{equation}
In \eqref{eq:langevin_update} and \eqref{eq:langevin_update_aux}, $U_{x\mu}^{(n)}$ denotes the link connecting the lattice point $x$ with its nearest neighbor in the $\mu$ direction after the $n$-th update, $\lambda_a$ are the Gell-Mann matrices and $\eps$ denotes the discrete Langevin-time step size. Moreover, $K_{x\mu a}$, the so-called drift term, when employing staggered fermions, reads
\begin{align}\label{eq:drift_term}
	\begin{aligned}
		K_{x\mu a}[U]& = \\
			-D_{x\mu a}&S_g[U]
			+\frac{1}{4}\sum_{f=u,d,s,c}\tr\left[M_f^{-1}[U]D_{x\mu a} M_f[U]\right]\;,
	\end{aligned}
\end{align}
where $S_g$ denotes the gauge action, $M_f$ is the Dirac matrix for the fermion flavor $f$, and we consider $2+1+1$ flavors. The trace in \eqref{eq:drift_term}, involving the left derivative
\begin{equation}
	D_{x\mu a}f[U] = \partial_\alpha f\left[e^{\ii\alpha\lambda_a}U_{x\mu}\right]_{\alpha=0}\;,
\end{equation}
is evaluated stochastically \cite{Sex14}. Finally, the $\eta_{x\mu a}^{(n)}$ in \eqref{eq:langevin_update_aux} denote independent real Gaussian noise variables with zero mean and unit variance. We note that the prescription \eqref{eq:langevin_update}, which was introduced in \cite{UF85}, is partially $\mathcal{O}(\eps)$ improved. This, together with our choice of keeping $\eps\leq10^{-4}$ in the employed adaptive-step-size algorithm \cite{AJS10}, ensures that finite-Langevin-step-size effects are smaller than our statistical errors, as confirmed by comparisons to hybrid Monte-Carlo (HMC) simulations at zero chemical potential.

We use a tree-level improved Symanzik gauge action and four steps of stout-smearing \cite{MP04} (generalized to $\SL(3,\C)$ in \cite{Sex19}) with a smearing parameter $\rho=0.125$. Denoting the numbers of lattice points in the spatial and temporal directions as $N_s$ and $N_t$, respectively, our lattices have an aspect ratio of $N_s/N_t=2$ and we consider $N_t=8$, $10$, $12$, and $16$. We change the  temperature via the inverse gauge coupling $\beta$ and work at physical pion masses. Indeed, our setup follows the one used in \cite{BBF15}, from which we also take the line of constant physics (LCP) along which we perform the continuum extrapolation $1/N_t^2 \to 0$ (omitting $N_t=8$). As a consistency check, we have performed vacuum runs using complex Langevin dynamics with bare parameters set by the LCP, computed the lattice spacing via the Wilson flow \cite{Lue10,BDF12} and found good agreement with the expected values. While our lattice volumes are smaller than those in \cite{BBF15}, we observe a negligible volume dependence in the measured observables, even for the highest temperatures, as confirmed by simulations with $N_t=10$ and aspect ratios up to $N_s/N_t=4$.

In principle, each quark flavor $f$ comes with its own chemical potential $\mu_f$. However, it is more common to work in a basis of conserved charges, which comprise the (net) baryon number $B$, electric charge $Q$, strangeness $S$, and charm $C$. The associated chemical potentials are related to the fundamental ones via
\begin{align}
	\begin{aligned}
		\mu_u = \frac{1}{3}\mu_B + \frac{2}{3}\mu_Q\;, \quad&
		\mu_d = \frac{1}{3}\mu_B - \frac{1}{3}\mu_Q\;, \\
		\mu_s = \frac{1}{3}\mu_B - \frac{1}{3}\mu_Q - \mu_S\;,\quad&
		\mu_c = \frac{1}{3}\mu_B + \frac{2}{3}\mu_Q + \mu_C\;.
	\end{aligned}
\end{align}
In our simulations, we set $\mu_s=\mu_c=0$ and $\mu_u=\mu_d=:\mu_l$, which implies $\mu_Q=0$ and corresponds to an isospin-symmetric setting. We vary only $\mu_l$ throughout our simulations. Although this setup is not exactly equivalent to the strangeness- and charm-neutral case, corresponding to the phenomenologically relevant conditions $\langle S\rangle=\langle C\rangle=0$ in the context of heavy-ion collisions (see \cite{BGK22}), it provides a very good approximation in the temperature regime considered here ($T \geq 260\,\mathrm{MeV}$). We mostly consider dimensionless quantities, normalized by appropriate powers of the temperature.

To enable stable simulations, we perform $16$ gauge-cooling steps \cite{SSS13} after every Langevin update \eqref{eq:langevin_update}. Moreover, in order to ensure that the stochastic evolution in our simulations produces correct results after equilibration, we carefully monitor the Langevin-time-evolution of the unitarity norm
\begin{equation}
	\mathcal{N}_U=\frac{1}{4\Omega}\sum_{x,\mu}\tr\left[\left(U_{x\mu}U^\dagger_{x\mu}-\id\right)^2\right]\;,
\end{equation}
with $\Omega=N_s^3N_t$. In particular, we restrict our ensembles to those configurations with $\mathcal{N}_U<0.1$, which has proven effective \cite{SSS20_2,HS22p} in avoiding the emergence of boundary terms or slowly decaying distributions that would spoil correct convergence \cite{ASS10,AJS11}. Finally, we do not expect problems related to the zeroes of the fermion determinant, corresponding to singularities in the drift term. This is because we are interested in high temperatures, for which it has been demonstrated that Dirac eigenvalues close to zero are generally absent \cite{ASS17,SSS20_2}. Hence, we are led to believe that our results should be unaffected by the wrong-convergence problem.

In our simulations, we scan the high-temperature region of the $(\mu_B,T)$ QCD phase diagram, restricting to temperatures above the crossover. We have generated $\mathcal{O}(10^2-10^3)$ configurations per simulation, usually combining two independent runs into one for better statistics, but ensuring proper thermalization for each. Measurements are taken in steps of $\tau=0.1$ Langevin time and remaining autocorrelations are handled via a conventional binned jackknife resampling, from which we estimate the statistical uncertainties. The bins are chosen such that they contain at least several integrated autocorrelation times. Our data \cite{data} and analysis scripts \cite{scripts} are available online.

	\emph{Results.}
	In the following, we discuss our results for the QCD equation of state at high temperatures and a broad range of densities. A comparable study using the same lattice setup in more conventional HMC simulations, circumventing the sign problem via a resummed Taylor expansion, was presented in \cite{BGK22}. While \cite{BGK22} considers lower temperatures and is limited to much smaller chemical potentials, we provide comparisons  wherever possible, i.e., for our lowest temperatures and small $\mu_B/T$.

The first observable of interest is the baryon density $n_B=\langle B\rangle/V$, with $V$ representing the lattice volume in physical units. Denoting the QCD partition function as $\mathcal{Z}$, $n_B$ is given by
\begin{equation}
	n_B = \frac{T}{V}\frac{\partial\ln\mathcal{Z}}{\partial\mu_B} = 
	\frac{T}{12V}\sum_{f=u,d,s,c}\left\langle\tr\left(M_f^{-1}\frac{\partial M_f}{\partial\mu_f}\right)\right\rangle\;.
\end{equation}
Similarly, one may define number densities for each quark flavor $f$,
\begin{equation}
	n_f = \frac{T}{V}\frac{\partial\ln\mathcal{Z}}{\partial\mu_f}\;.
\end{equation}
With this notation, we find that $n_B=(n_u+n_d)/3 = n_l/3$, since $n_s$ and $n_c$ vanish to numerical precision, consistent with the approximate strangeness and charm neutrality.

\begin{figure}[t]
 	\centering
	\includegraphics[scale=0.68]{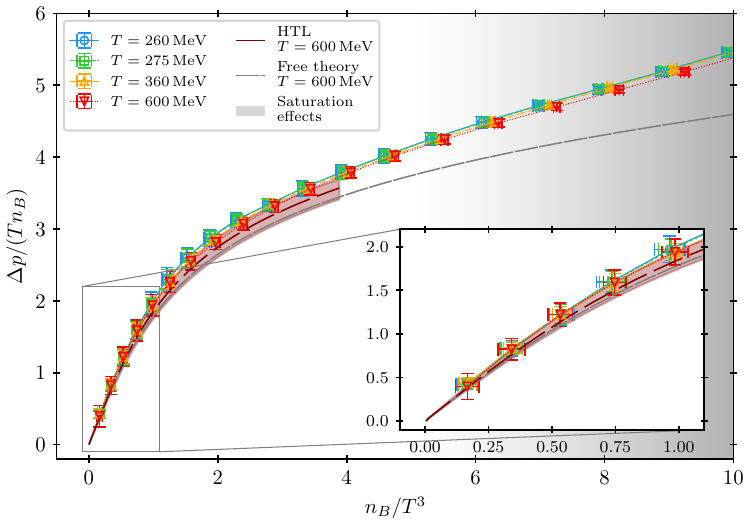}
	\caption{Pressure difference (normalized by $n_B$) as a function of $n_B/T^3$ for different temperatures $T$. HTL and free-theory results are shown for comparison.}
 	\label{fig:deltaP_vs_nB}
\end{figure}

We have computed $n_B$ for various values of $\mu_B$ and $T$ and for different $N_t$. For fixed $T$ and $N_t$, we interpolate the $\mu_B$ dependence of $n_B$ and extrapolate these interpolations to the continuum limit along the LCP. Statistical uncertainties are calculated using a jackknife routine, as described above. Moreover, systematic errors are estimated by varying several aspects of the analysis: the interpolation method in the $\mu_B$ direction (a cubic spline and a polynomial of ninth degree containing only odd powers to respect charge conjugation symmetry), the subset of data points used for the interpolation (all, even, or odd), the number of samples omitted to account  for thermalization, and the number of bins in the jackknife routine. This procedure yields a total of $96$ different results, the median of which is quoted as the final value. Furthermore, the statistical uncertainty is set to the maximum of the $96$ statistical uncertainties, while the systematic error is estimated as half the width of the central $68\%$ percentile interval. The final error estimate is taken to be the sum of the statistical and systematic uncertainties. \footnote{Throughout, we indicate the combined statistical and systematic uncertainties of our results as error bands in the figures. The exception is \cref{fig:deltaP_vs_nB}, in which we show error bars instead.}

\begin{figure}[t]
	\centering
	\includegraphics[scale=0.68]{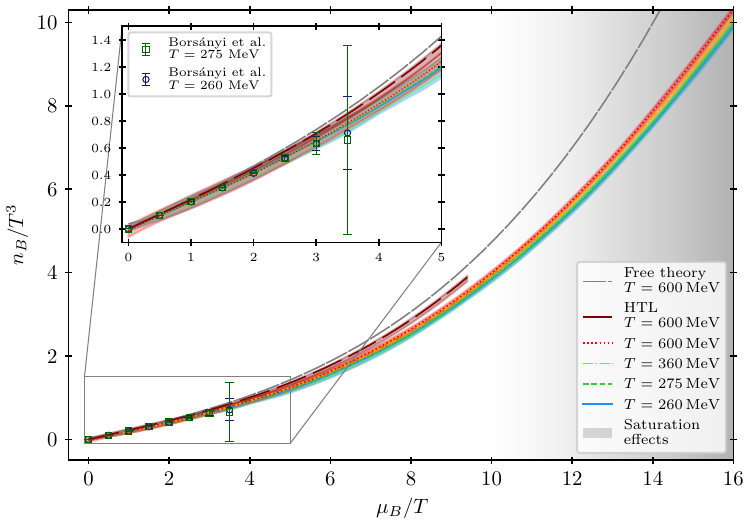}
	\caption{Continuum-extrapolated baryon density $n_B$ as a function of the baryon chemical potential $\mu_B$ (both normalized by powers of $T$) for different temperatures $T$. Comparisons with \cite{BGK22} and HTL results are shown for small $\mu_B$. Moreover, the gray dashed line corresponds to a free theory of quarks and gluons, while the inset shows a close-up view of the small-$\mu_B$ region. Finally, the shaded region indicates the onset of saturation effects, see the main text.}
	\label{fig:density_vs_mu}
\end{figure}

The result of this procedure for different temperatures is shown in \cref{fig:density_vs_mu}. The normalized quantity $n_B/T^3$ shows the expected increase with $\mu_B/T$ and a negligible $T$ dependence if $\mu_B/T$ is kept fixed. As indicated in the plot, our results are consistent with the results of \cite{BGK22} within the region of the phase diagram that is accessible to both studies. In addition, for small $\mu_B/T$ we find agreement with hard-thermal-loop (HTL) perturbative results derived in \cite{HAM14,HBA14,AHM16}, as well as with the baryon density of a theory of noninteracting quarks and gluons, only the former of which contribute to $n_B$. It should be noted that the HTL results consider only three quark flavors and we have na\"ively extended them up to $\mu_B\approx3\pi T$, beyond which they are no longer expected to be valid according to \cite{HBA14} due to the involved expansions in $\mu_B/T$. Notably, however, our results start to deviate from the HTL baryon density (as well as from the free-theory result) already for $\mu_B$ smaller than that bound.

Since we study a finite, lattice-discretized system, only a finite number of fermionic modes is available. Thus, at sufficiently large chemical potentials, all modes become occupied and the baryon density ceases to increase. This saturation effect becomes relevant at a chemical potential value that depends on the UV cutoff. The onset and severity of saturation effects are indicated by the shaded region in \cref{fig:density_vs_mu}. More precisely, we define the onset of saturation as the chemical potential beyond which our continuum extrapolation becomes unreliable according to a $\chi^2$ test. In practice, this onset corresponds to $n_l/n_{l,\mathrm{sat}} \approx 0.02$ on our coarsest lattice ($N_t=10$), where $n_{l,\mathrm{sat}}=3/2$ (in lattice units) denotes the saturation density. At the same chemical potential, we find $n_l/n_{l,\mathrm{sat}} \approx 0.003$ on our finest lattice ($N_t=16$). Thus, while on the finest lattice saturation effects are likely very small, the same cannot be said for the coarsest lattice, which causes the continuum extrapolation to become unreliable. The $\chi^2$ value increases with $\mu_B/T$, indicating that saturation effects become increasingly more severe at larger $\mu_B/T$.

With the baryon density, we can now compute the pressure $p=\frac{T}{V}\ln\mathcal{Z}$, the quantity of central importance for the equation of state. The pressure difference with respect to vanishing chemical potential,
\begin{equation}
	\Delta p(T,\mu_B)=p(T,\mu_B)-p(T,0)\;,
\end{equation}
can be obtained via \cite{Sex19}
\begin{equation}\label{eq:pressure}
	\frac{\Delta p(T,\mu_B)}{T^4}=\frac{1}{T^4}\int_0^{\mu_B} d\mu_B'n_B(T,\mu_B')\;,
\end{equation}
where the integral is computed by numerically integrating the aforementioned interpolations of $n_B$. The zero chemical potential contribution $p(T,0)$ is taken as input from (the Supplementary Information of) \cite{BFG16}. Note that the uncertainties of those data contribute to the total uncertainty.

\begin{figure}[t]
	\centering
	\includegraphics[scale=0.68]{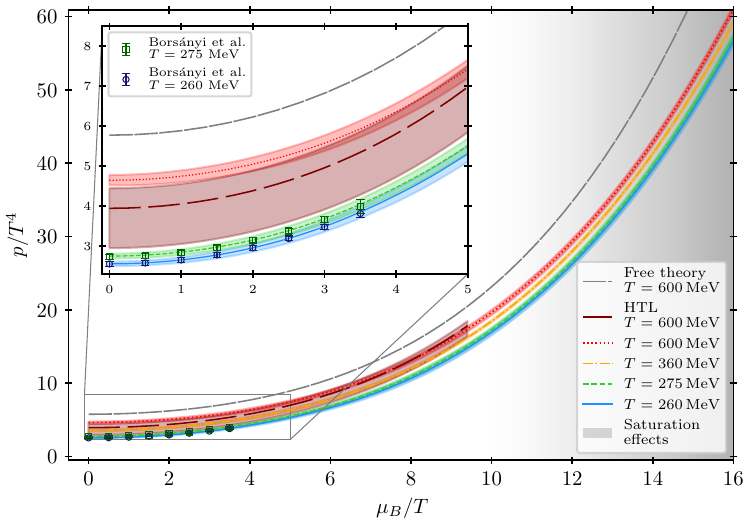}
	\caption{Similar as \cref{fig:density_vs_mu}, but for the pressure.}
	\label{fig:pressure_vs_mu}
\end{figure}

The continuum-extrapolated result for the pressure as a function of $\mu_B$ is shown in \cref{fig:pressure_vs_mu}. Notably, the ($T$-normalized) pressure shows only a weak temperature dependence for fixed $\mu_B/T$. We once again compare with \cite{BGK22} and find no disagreement. Furthermore, as before we provide comparisons with HTL results and observe agreement for all $\mu_B/T$ for which the HTL expansion is believed to be valid \cite{HBA14}. Finally, our results deviate from the pressure of a free theory for all chemical potentials considered. Such a deviation is present already for $p(T,0)$ and is not unexpected for these temperatures \cite{KLP00}.

Being the central quantity discussed in this work, we have fitted the $\mu_B$ and $T$ dependence of the pressure with a polynomial ansatz. Within the temperature range $260\,$MeV$\leq T\leq600\,$MeV and up to $\mu_B\leq10\,T$, our results are consistent with
\begin{align}\label{eq:pressure_fit}
	\begin{aligned}
		\frac{p(T,\frac{\mu_B}{T})}{T^4} =& \\ \sum_{i=0}^8c_i&\left(\frac{T}{T_\mathrm{ref}}\right)^i + \sum_{j=0}^2\sum_{k=0}^2 c_{jk}\left(\frac{T}{T_\mathrm{ref}}\right)^j\left(\frac{\mu_B}{T}\right)^{2(k+1)}\;,
	\end{aligned}
\end{align}
where $T_\mathrm{ref}=300\,$MeV is a reference temperature and the fit coefficients are given by
\begin{align}\label{eq:pressure_fit_coefficients}
	\begin{aligned}
	&c_0 = -6.8283\times10^0\;, \quad
	&&c_1 = 2.0917\times10^1\;,\\
	&c_2 = -1.5003\times10^1\;, \quad
	&&c_3 = 1.7108\times10^0\;, \\
	&c_4 = 4.9476\times10^0\;, \quad
	&&c_5 = -3.7892\times10^0\;, \\
	&c_6 = 1.2770\times10^0\;, \quad
	&&c_7 = -2.1378\times10^{-1}\;, \\
	&c_8 = 1.4440\times10^{-2}\;, \quad
	&&c_{00} = 9.7969\times10^{-2}\;, \\
	&c_{01} = -1.7212\times10^{-4}\;, \quad
	&&c_{02} = 4.0490\times10^{-6}\;,\\
	&c_{10} = 4.5298\times10^{-3}\;, \quad
	&&c_{11} = 8.3163\times10^{-4}\;,\\
	&c_{12} = -5.1857\times10^{-6}\;, \quad
	&&c_{20} = -2.1202\times10^{-3}\;,\\
	&c_{21} = -2.4106\times10^{-4}\;, \quad
	&&c_{22} = 1.5487\times10^{-6}\;.
	\end{aligned}
\end{align}
Over the range of temperatures and chemical potentials considered above, this parametrization remains consistent with the numerical results within their uncertainties. Consequently, these uncertainties can be carried over to the analytical formula \eqref{eq:pressure_fit}, amounting to at most 5\% deviation. From these results, the remaining thermodynamic observables could be derived straightforwardly. Hence, \eqref{eq:pressure_fit} and \eqref{eq:pressure_fit_coefficients} can be considered the central result of this work.

\begin{figure}[t]
 	\centering
	\includegraphics[scale=0.68]{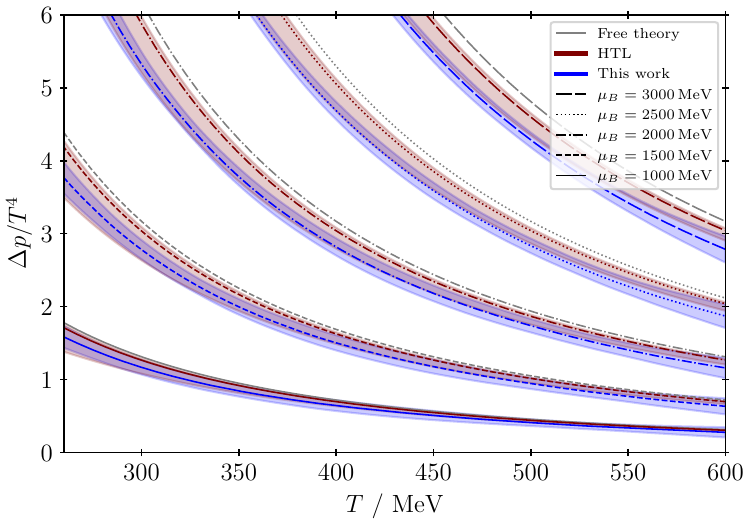}
	\caption{Pressure difference $\Delta p$ as a function of temperature $T$ for different chemical potentials $\mu_B$ and a comparison to HTL results and a free theory of quarks and gluons.}
 	\label{fig:deltaP_vs_T}
\end{figure}

In \cite{HBA14}, HTL results for the pressure difference $\Delta p$ as a function of $T$ were compared with (older) lattice data for different small chemical potentials and good agreement was found. A similar comparison of our results for $\Delta p$ with HTL perturbation theory and a free-theory calculation but for larger chemical potentials is shown in \cref{fig:deltaP_vs_T}. Once more, one observes reasonable agreement between the results for small $\mu_B/T$ and larger deviations upon increasing the chemical potential. Rather consistently, the HTL results lie between the free theory and our nonperturbative results.

\begin{figure}[t]
 	\centering
	\includegraphics[scale=0.68]{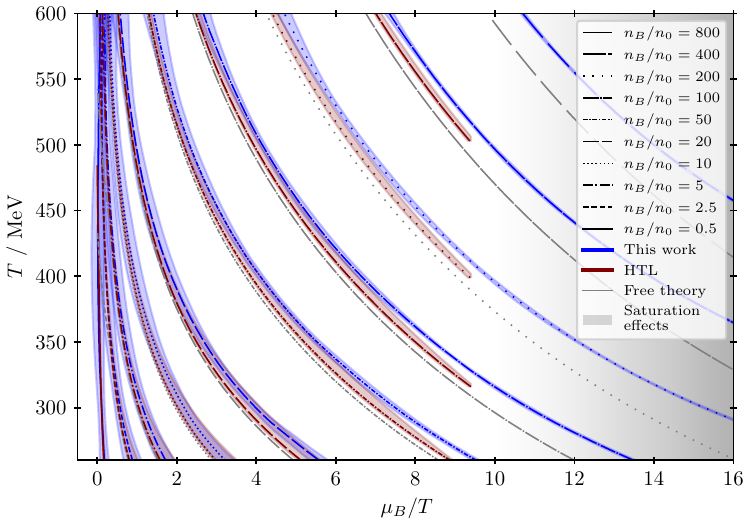}
	\caption{Lines of constant baryon density $n_B$ (normalized by the nuclear density $n_0$) as a function of both $\mu_B/T$ and $T$. HTL and free-theory results are shown for comparison.}
 	\label{fig:lines_of_constant_density}
\end{figure}

Lines of constant baryon density $n_B$ (normalized by the nuclear density $n_0=0.16\,\mathrm{fm}^{-3}$) in the $(\mu_B/T,T)$ plane are shown in \cref{fig:lines_of_constant_density}. As can be seen, we reach baryon densities as high as $n_B\gtrsim400\,n_0$ in our simulations. Finally, in \cref{fig:deltaP_vs_nB}, we show the pressure difference normalized by the baryon density as a function of $n_B$. In the low-density region, we observe the linear increase of $\Delta p/(Tn_B)$ with $n_B/T^3$ expected from a leading-order expansion. This behavior, however, changes upon further increasing the density due to the contribution of higher-order terms.

	\emph{Conclusions.}
	We have presented the first complex Langevin simulations of high-temperature QCD at the physical point, including a continuum extrapolation, reaching baryon densities that far surpass the state of the art of conventional approaches based on importance sampling. Finite-volume effects are under control. In particular, we have computed the pressure for a broad range of temperatures (above the QCD crossover) and baryon densities. We have not encountered any symptoms of the wrong-convergence problem, which lends further support to our results. Our study therefore represents a significant advancement over previous work and may be useful, for instance, in cosmological applications and in modeling heavy-ion collisions.

A logical next step would be to extend our simulations to lower temperatures. This, however, would come with excessive numerical costs due to convergence issues in the conjugate gradient algorithm employed for matrix inversion. Moreover, problems due to the poles in the drift term caused by small Dirac eigenvalues \cite{ASS17,NS15} could no longer be excluded. Without further (algorithmic or methodological) improvements, such simulations are currently unfeasible. We mention, however, current efforts \cite{MSS26p} of machine-learning so-called kernels that could stabilize simulations in the low-temperature regime of QCD like they have in other systems \cite{LS23,ALR23,ARS24,BHM23,BHM24_2}. Further interesting research directions include the addition of electromagnetic fields \cite{BBE14,EM24} and/or an isospin chemical potential \cite{BCE23}.

\vspace{0.3cm}
\emph{Acknowledgments.}
We would like to thank Szabolcs Bors\'anyi and Jana G\"unther for providing the HMC results used above as well as data for the LCP. Moreover, we thank Najmul Haque for sending us \verb|Wolfram Mathematica| notebooks to produce the HTL data used for comparison in this work. We are indebted to Astrid Carstensen, Erhard Seiler, and Ion-Olimpiu Stamatescu for collaborations on related topics. The computational results have been achieved using the Austrian Scientific Computing (ASC) infrastructure as well as the computing cluster of the university of Graz (GSC). This research was funded in whole or in part by the Austrian Science Fund (FWF) \href{https://doi.org/10.55776/P36875}{10.55776/P36875}. The data analysis performed in this work is based on the \verb|Python| ecosystem for scientific computing \cite{python}, in particular via the packages \cite{numpy,matplotlib,scipy,pandas} and we are also grateful for the creation and maintenance of all their dependencies.

\emph{Open Access Statement.}
For the purpose of open access, the author has applied a CC BY public copyright license to any Author Accepted Manuscript version arising from this submission.

\emph{Data Availability Statement.}
The data set underlying this work \cite{data} as well as the employed analysis scripts \cite{scripts} are published online and our simulation code is available upon request.

\bibliographystyle{apsrev4-2}
\bibliography{bibliography}

\end{document}